\documentclass[aps,prl,twocolumn,showpacs,showkeys]{revtex4}

\usepackage{graphicx}
\usepackage{amsfonts}
\usepackage{amsmath}
\usepackage{amssymb}
\usepackage{amsfonts}
\usepackage{graphicx}

\begin{document}
\bibliographystyle{unsrt}

\title{Current Induced Excitations in Cu/Co/Cu Single Ferromagnetic
Layer Nanopillars }
\author{B. \"{O}zyilmaz and A. D. Kent}
\address{Department of Physics, New York University, New York, NY 10003, USA}
\author{J. Z. Sun, M. J. Rooks and R. H. Koch}
\address{IBM T. J. Watson Research Center, P.O. Box 218, Yorktown
Heights, NY 10598, USA}

\begin{abstract}
Current-induced magnetic excitations in Cu/Co/Cu single layer
nanopillars ($\sim$50 nm in diameter) have been studied
experimentally as a function of Co layer thickness at low
temperatures for large applied fields perpendicular to the layers.
For asymmetric junctions current induced excitations are observed
at high current densities for only one polarity of the current and
are absent at the same current densities in symmetric junctions.
These observations confirm recent predictions of spin-transfer
torque induced spin wave excitations in single layer junctions
with a strong asymmetry in the spin accumulation in the leads.
\end{abstract}
\volumeyear{year} \volumenumber{number} \issuenumber{number}
\eid{identifier}
\date{March 15, 2004}





\maketitle

Angular momentum transfer studies in magnetic nanostructures have
made tremendous progress during the last few years. Recently, both
spin current induced magnetization reversal
\cite{Katine,Grollier1,Barbaros-PRL} and spin current driven
magnetization precession \cite{Kiselev, Rippard-PRL} have been
directly observed in magnetic nanostructures. These experiments
confirmed seminal predictions by Berger \cite{Berger} and
Slonczewski \cite{Slonczewski1}, that a magnet acting as a
spin-filter on a traversing current can experience a net torque:
(spin-) angular momentum which is filtered out of the current must
be absorbed by the ferromagnet. In the presence of a significant
angular momentum component transverse to the magnetization of the
ferromagnet this leads to a so called spin-transfer torque. A
transverse spin-polarization of the electric current was thought
to be necessary for current induced excitations of the
magnetization. Hence most of the experimental and theoretical work
on spin-transfer torque concentrated on spin valve type structures
of ferromagnet/normal metal/ferromagnet layers, in which the layer
magnetizations may be non-collinear. Only recently, the necessity
of a transverse component of spin polarized current has been
relaxed \cite{Polianski,Stiles-Single}. At high enough current
densities Polianski \textit{et al.} \cite{Polianski} and Stiles
\textit{et al.} \cite{Stiles-Single} predict spin wave excitations
in thin ferromagnetic layers even when the current is unpolarized.

Polianski \textit{et al.} \cite{Polianski} have reemphasized the
spin-filtering property of a ferromagnet (FM) as the fundamental
cause for spin transfer torque. Spin-filtering is present also in
normal metal/ferromagnetic metal/ normal metal (NM/FM/NM) pillar
junctions with only a single FM layer. In the current
perpendicular to the plane geometry a current bias results in spin
accumulation on either side of the FM. Fluctuations in the
magnetization direction combined with spin diffusion
parallel to the NM/FM interfaces result in a spin-transfer torque.
At each interface these torques act to align the magnetization
along the direction of the spin accumulation. In a perfectly
symmetric single layer structure the resulting torques are of
equal magnitude but opposite direction and cancel each other.
However, if the mirror symmetry is broken the torques acting on
each NM/FM interface have different magnitudes. For this case,
Ref. \cite{Polianski,Stiles-Single} predict that an unpolarized
current can induce spin wave instabilities and generate spin-wave
excitations with wavevectors in the film plane. Instabilities
occur when the current bias is such that the direction of the
larger spin accumulation is anti-parallel to the direction of the
magnetization of the FM. Polianski \textit{et al.}
\cite{Polianski} studied the case of an thin FM with the
magnetization being fixed along the current flow direction. Here,
the break in symmetry requires asymmetric contacts. Stiles
\textit{et al}. \cite{Stiles-Single} relaxed this requirement and
allowed the magnetization to vary along the current flow
direction, which also breaks the mirror symmetry. In either case
in ideal asymmetric junctions current induced excitations are
predicted to occur for only one current polarity and are expected
to be absent in perfectly symmetric structures. Both groups made
predictions on how single layer instabilities depend on parameters
such as the current bias polarity, the FM layer thickness, the
degree of asymmetry of the single layer junction and the applied
field.

In this letter we report systematic studies of current induced
excitations of the magnetization in both symmetric and asymmetric
nanopillar junctions containing only a single FM layer.
Measurements were performed in high magnetic fields ($H > 4\pi M$)
in the field perpendicular to the plane geometry at 4.2 K. For
sufficiently large current densities we observe anomalies in
$dV/dI$ for only one current polarity. Current induced single
layer excitations occur only in asymmetric pillar devices (PD) and
lead to a decrease of the junction resistance. They are absent in
symmetric PDs. Our results confirm the recent prediction of
current induced excitations in asymmetric PDs.

Pillar junctions $\sim$50 nm in size have been fabricated by means
of a nano-stencil mask process \cite{Sun-APL}, which has been used
earlier for spin-transfer torque studies in Co/Cu/Co trilayer spin
valves \cite{Sun3, Barbaros-PRL}. To study the thickness
dependence of single layer excitations we combined the
nano-stencil mask process with an \textit{in-situ} wedge growth
mechanism. With this approach we have fabricated PDs with a single
Co layer of continuously varied thickness across a single wafer
\cite{Barbaros- to be published}. As shown in Fig. 1, structures
fabricated by means of an undercut template are intrinsically
asymmetric due to the requirement of an inert bottom electrode
surface, usually Pt, on top of which the pillar structure is
grown. Here, asymmetry refers to the spin-accumulation pattern
generated within the PD with respect to the Co layer position. The
strong asymmetry due to the choice of Pt as bottom electrode is
removed by inserting a second Pt layer. Therefore, the study of
spin-transfer in symmetric single layer structures requires the
``capping" of the pillar with a Pt layer as indicated in Fig. 1.
Many junctions with a FM layer thickness varying from 2 nm to 17
nm and lateral dimensions from 30 nm $\times$ 60 nm up to 70 nm
$\times$ 140 nm have been studied as a function of bias current
and applied field. The range of Co layer thickness covers both the
case where the thickness $t$ is smaller than the exchange length
$l_{ex}$ of Co and the case where the thickness is comparable to
the latter ($t \geq l_{ex}$). All junctions in this thickness
range exhibit single layer excitations. Here we discuss
representative data obtained on PDs with $t\approx$8 nm and
$t\approx$17 nm and lateral dimensions of 30 nm $\times$ 60 nm and
50 nm $\times$ 50 nm respectively. To confirm that the excitations
are caused by asymmetric contacts we have repeated experiments
with symmetric PDs with a stack sequence of $|$PtRh15 nm$|$Cu10
nm$|$Co10 nm$|$Cu10 nm$|$Pt 15nm$|$.

All measurements reported here were conducted at 4.2 K in a four
point-geometry configuration in fields applied perpendicular to
the thin film planes. The differential resistance dV/dI was
measured by lock-in technique with a 100 $\mu$A modulation current
at $f=873$ Hz  added to a dc bias current. As shown in Fig. 1
positive current is defined such that the electrons flow from the
bottom electrode of the junction to the top electrode.

A typical magnetoresistance (MR) measurement of a single layer
junction at 0 dc bias is shown in Fig. 1.
\begin{figure}
\begin{center}\includegraphics[width=8.5cm]{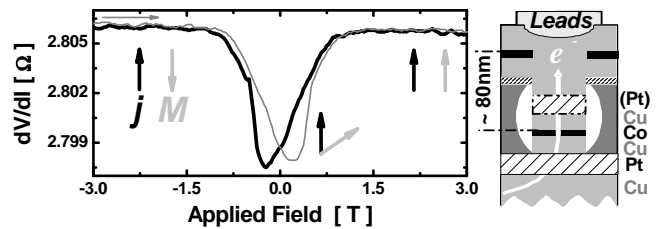}
\vspace{-2 mm}\caption{Left: typical $dV/dI$ vs $H$ measurement at
0 DC bias. The junction size is 50 nm $\times$ 50 nm and
$t\approx$17 nm. An increase in junction resistance $(\sim
0.1$\%$)$ is observed when $\mathbf{j}$ and $\mathbf{M}$ are
collinear. Right: schematic of a single Co layer pillar junction
fabricated via the nano-stencil mask process. Electron flow
indicates the definition of positive current bias. Symmetric
junctions are fabricated by addition of a Pt layer (dash-dotted
box). }\vspace{-6 mm}
\end{center}
\end{figure}
The resistance $R$ has its minimum when the magnetization
$\mathbf{M}$ lies in the thin film plane, i.e. when $\mathbf{M}$
is orthogonal to $\hat{\mathbf{j}}$. We observe a gradual increase
in $R$ as we increase the applied field which tilts the
magnetization vector out of the thin film plane. Once the applied
field exceeds $4\pi M$, $\mathbf{M}$ is collinear with
$\hat{\mathbf{j}}$ and the resistance saturates at its maximum.
From this we conclude that the observed MR is sensitive enough to
register (\textit{field induced}) changes of relative orientation
of $\hat{\mathbf{j}}$ and $\mathbf{M}$. This provides a convenient
$``in-situ"$ tool for detecting also \textit{current induced}
changes of the magnetization. It is important to note, that for
even the thickest layer we observe a \textit{decrease} of the
resistance in the field sweeps when $\mathbf{M}$ and
$\hat{\mathbf{j}}$ start deviating from collinear alignment.

A typical I(V) curve for an asymmetric single layer PD is shown in
Fig. 2(a). Here $dV/dI$ versus $I$ is plotted for fields $H = 1.5$
T, 2 T, 2.5 T and $H = 3.1$ T for a 30 nm $\times$ 60 nm junction
with $t\approx8$ nm. At fields above the demagnetization field
($H>1.5$ T) we observe anomalies in the form of small dips at
negative current polarity only. The presence of many modes makes
it difficult not only to distinguish individual modes but also to
find the threshold current for single layer excitations at a
particular field value. Note that in the field perpendicular
geometry the onset of these excitations always leads to a (small)
decrease in resistance, which is opposite to what has been
observed in both point contact experiments \cite{RalphScience,
JiPRL, footnote1} and trilayer PDs.

To distinguish these excitations from the parabolic background
resistance, we plot $d^{2}V/dI^{2}$, which is sensitive to abrupt
features in $dV/dI$. Plotted on a greyscale as a function of the
applied field and the current bias it represents a phase diagram
for single layer excitations. An example of such a plot is shown
in Fig. 2(c). Here the current is swept from -15 mA to +15 mA
while the magnetic field is held constant for each current sweep.
For subsequent sweeps the field is stepped from -4.6 T to +4.6 T
in 100 mT steps. The ``current bias-applied field" plane
segregates into two regions separated by a straight line, which we
associate with the threshold current, the critical current
$I_{crit}$ for single layer excitations. For fields $H>4\pi M$
excitations  only occur for negative current polarities. At
negative current bias excitations are absent below the critical
current, whereas above the current threshold many modes are
excited. $I_{crit}$ shows a linear dependence on the applied field
and can be extrapolated approximately to the origin. Dividing
$I_{crit}$ by the \textit{nominal} junction area $A$, we estimate
the field dependence of the critical current density $j_{crit}=bH$
with $b\approx1.9\times10^{8}$ (A/cm$^{2}$)/T. We obtain a more
accurate estimate for $j_{crit}$ by multiplying $I_{crit}$ with
the junction resistance $R\approx 2.55 \Omega$ , which is
equivalent to dividing by an \textit{effective} junction area:
$j_{crit}\propto I_{crit} R=\beta H$ with $\beta \approx 8.8
\times 10^{-3}$ (A$\Omega$/T).

A better way to distinguish the small features of current induced
excitations from the varying background resistance is to fix the
latter. This can be done by keeping the current constant and
sweeping the applied field instead. Here an example of such a
measurement is given in Fig 3(a) and (c). Field sweeps at fixed
negative current bias are shown in Fig. 3(a), whereas Fig. 3(c)
shows the MR at fixed positive currents. The strongest evidence
for current induced excitations in single layer junctions comes
from the comparison of these two figures. As shown in Fig. 3(c)
excitations at fields $H>4\pi M$ are absent in the field traces.
However, high current densities at positive bias gradually
increase the applied field at which the differential resistance
saturates.
\begin{figure}
\begin{center}\includegraphics[width=8.5cm]{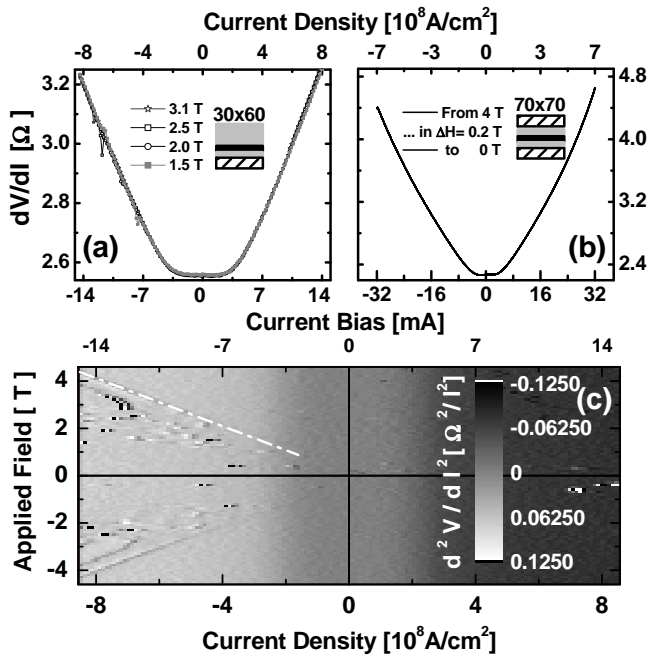}
\vspace{-2 mm}\caption{ $dV/dI$ vs $I$ at constant fields. (a)
asymmetric junction (30 nm$ \times$ 60 nm, $t\approx8$ nm) with Pt
as bottom electrode. For $H>4\pi M$ dips are observed at negative bias only.
(b) Symmetric junction (70 nm $\times$ 70 nm, $t\approx10$ nm)
with Pt on either side of the Co layer ($t\approx10$ nm). I(V)
curves at different field values overlap fully. (c) Phase diagram
for current induced excitations in single layer junctions; same
junction as in Fig. 2(a). $d^{2}V/dI^{2}$ is plotted on a grayscale.
The white dash-dotted line indicates the boundary
for excitations. }\vspace{-6 mm}
\end{center}
\end{figure}
This effect cannot be attributed solely to the presence of
additional (Oersted) fields related to the charge current and is
not yet fully understood. There is a dramatic change in the field
traces if one applies a negative current bias to the junction. For
each fixed current value there is now a critical field $H_{crit}$,
above which the resistance remains constant. However, below
$H_{crit}$ the observation of peaks and dips indicates the
presence of many (current induced) excitations. $H_{crit}$ is a
linear function of the bias current and shifts to higher values as
one increases the current. As can be seen in Fig. 3(b), the linear
fit of the critical fields can once more be extrapolated to the
origin. Hence in both field sweeps at fixed currents and current
sweeps at fixed fields one obtains a linear dependence of the
critical parameter on the running variable, i.e $j_{crit}=bH$ and
$H_{crit}=cj$. For a particular Co layer thickness the slopes $b$
and $c$ are equivalent, i.e. $b \cong c^{\textrm{-1}}$. From Fig.
3(b) and the nominal junction area $A$ we estimate the current
density dependence of $H_{crit}=cj$ with
$c\approx5.2\times10^{-9}$ T/(A/cm$^{2}$). Using the junction
resistance $R\approx2.80\Omega$ as an approximation for the
effective junction area we obtain $H_{crit}\propto \zeta(I R)$
with $\zeta\approx$ 73.8 T/(A$\Omega$). Note that for $H<4\pi M$
there are large changes in the hysteresis for both current
polarities. We attribute these changes to the interaction of the
Oersted fields with magnetic domain configurations \cite{Barbaros-
to be published}.

We have also studied the thickness dependence of these excitations
and summarize the results in Fig. 3(d). For all thicknesses the
observed boundary in the ``current bias/applied field plane" can
be extrapolated close to the origin. Here we only plot the slope
$\beta$ of the field dependence of $I_{crit} R$ ($\propto
j_{crit})$ as a function of Co layer thickness $t$. We observe an
\textit{ increase} of $\beta$ with increasing $t$,
$\Delta\beta/\Delta t \approx (0.48 \pm 0.05)$ (mA$\Omega$)/(T
nm). The critical currents increase by approximately a factor of
two as one increases the Co layer thickness $t$ from 2 nm to 17
nm. Over the same thickness range the junction resistance $R$
increase only by $\approx25$\% (not shown).

To clarify the origin of these excitations, we have repeated these
experiments in symmetric single layer PDs. An example of current
sweeps at fixed fields in these structures is shown in Fig. 2(b).
Here the current is swept from +32 mA to -32 mA in a 70
  nm $\times$ 70 nm junction. In magnetic fields up to 4 T features
such as dips or peaks are absent in the current-voltage
characteristics. Also, field sweeps at fixed current do not
exhibit any of the strong polarity dependence observed in
asymmetric PDs. To summarize, in symmetric junctions current
induced excitations are absent up to $j\leq 7\times10^{8}$
A/cm$^{2}$ .

Experimental results and theoretical predictions are in good
agreement. Both models give the correct order of magnitude,
correct polarity  and thickness dependence of $j_{crit}$ in
asymmetric structures. Ref. \cite{Polianski} studied the case of
uniform magnetization $\mathbf{M}$ in the current flow direction
$\hat{\mathbf{j}}$. Ref. \cite{Stiles-Single} also considered the
case where  $\mathbf{M}$ is allowed to vary along
$\hat{\mathbf{j}}$. For this case excitations are expected to
occur independent of current polarity even in \textit{symmetric}
PDs. However, the predicted critical currents are much larger
($j_{crit} >10^{10}$ A/cm$^{2}$) than for the asymmetric case
\cite{Stiles private communications}. Once $\mathbf{M}$ is allowed
to vary along $\hat{\mathbf{j}}$, current induced excitations are
predicted for both current polarities, albeit, with large
differences in the magnitude of critical currents. For example for
an asymmetric junction with $t\approx17$nm the necessary
\textit{positive} current densities ($j_{crit}>2.5\times10^{9}$
A/cm$^{2}$) far exceed the value which can be sustained by
existing PDs. The linear dependence of $j_{crit}$ on $H$ can be
explained by both models. The (near) zero intercept of $j_{crit}$
is somewhat peculiar but can also be explained if the influence of
the shape and finite size of the PD on the spin wave modes is
properly accounted for in models \cite{Stiles private
communications}. Also the increase of the critical current
$j_{crit}$ with increasing Co layer thickness $t$ is in agreement
with theoretical predictions. An \textit{increase} of $j_{crit}$
with increasing $t$ is expected due to an increase of the (bulk)
damping \cite{Polianski, Stiles-Single}. According to Ref.
\cite{Stiles-Single} in thicker films ($t\gtrapprox l_{ex}$) the
variation of $\mathbf{M}$ along $\mathbf{\hat{j}}$ introduces an
additional source of asymmetry. This should activate a competing
effect which by itself would \textit{decrease} $j_{crit}$ with
increasing $t$. However, to determine which effect would dominate
details of layer structure and junction geometry need to be
considered. The direct comparison between experimental results and
theoretical predictions is further hampered by the change of
asymmetry in spin accumulation as we increase the Co layer
thickness \cite{footnote2}. For our device geometry and for Co
layer thicknesses up to $t \sim17$ nm ($t>l_{ex})$ the dominant
source of the current-induced excitations appears to be the asymmetry of
the leads.

\begin{figure}
\begin{center}
\includegraphics[width=8.5cm]{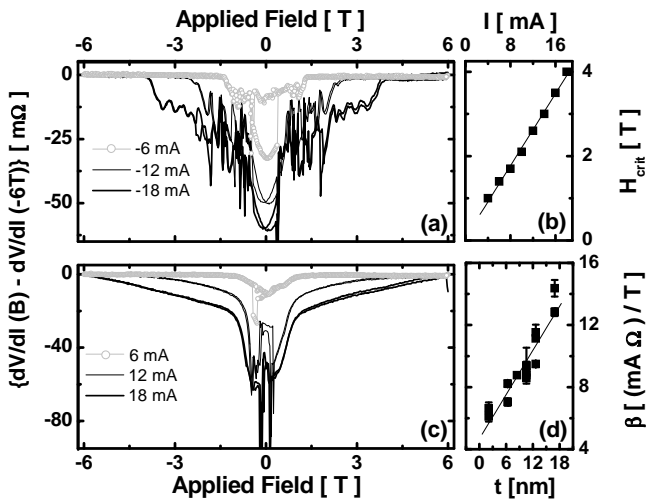}
\vspace{-3 mm} \caption{ (a) $dV/dI$ vs $H$ at negative current
bias. The zero dc bias field sweep of this junction is shown in
Fig. 1. (b) Current bias dependence of the critical fields above
which excitations are not observed. (c) $dV/dI$ vs $H$ for
positive current bias; excitations are absent. (d) Thickness
dependence of the ``critical currents." Here the slope $\beta$ of
$I_{crit} R$ is plotted as a function of Co layer thickness $t$.}
\vspace{-3 mm}
\end{center}
\end{figure}

Finally we would like to address the possibility of current
induced excitations in multilayered structures caused by an
asymmetry in spin accumulation in the leads. For trilayer
structures with a stack sequence of $|$Pt$|$Cu$|$Co
(thin)$|$Cu$|$Co (thick)$|$Cu$|$ parallel orientation of the
magnetization results in a spin accumulation asymmetry at the
thick layer similar to the one in single layer junctions discussed
above. Hence, high negative currents should lead to spin wave
instabilities. Also the anti-parallel configuration leads to a
strong asymmetry in spin accumulation at the thicker layer.
However, the asymmetry in spin accumulation at the interfaces of
the thick layer is now reversed. Therefore, spin wave
instabilities are now conceivable for positive current bias.
Consequently, a strong asymmetry in spin accumulation should lead
to spin wave instabilities in trilayer nanopillars for
\textit{both} current polarities at current densities, similar to
those at which magnetization reversal is observed.

In conclusion we have studied current induced spin wave
excitations in symmetric and asymmetric pillar junctions with only
a single ferromagnetic layer. We have confirmed that excitations
occur in asymmetric junctions and are absent in symmetric
junctions at similar current densities. We have also shown that in
asymmetric junctions the critical currents increase with Co layer
thickness. Finally, we have discussed implications of an asymmetry
in longitudinal spin accumulation in Co/Cu/Co trilayers.

\begin{acknowledgments}
This research is supported by grants from NSF-FRG-DMS-0201439
and by ONR N0014-02-1-0995.

\end{acknowledgments}



\begin{thebibliography}{99}
%
\bibitem{Katine} J. A. Katine \textit{et al.}, Phys. Rev. Lett.
\textbf{84}, 3149 (2000).
\bibitem{Grollier1} J. Grollier \textit{et al.}, Appl. Phys. Lett.
\textbf{78}, 3663, (2001).
\bibitem{Barbaros-PRL} B. \"{O}zyilmaz \textit{et al.}, Phys. Rev.
Lett. \textbf{91}, 067203 (2003).
\bibitem{Kiselev} S. I. Kiselev \textit{et al.}, Nature \textbf{425},
380 (2003).
\bibitem{Rippard-PRL} W. H. Rippard \textit{et al.}, Phys. Rev. Lett.
\textbf{92},
027201, (2003).
\bibitem{Berger} L. Berger, Phys. Rev. B \textbf{54}, 9353 (1996).
\bibitem{Slonczewski1} J. Slonczewski, J. Magn. Magn. Mater.
\textbf{159}, L1(1996).
\bibitem{Polianski} M. L. Polianski, P. W. Brouwer, Phys. Rev. Lett.
\textbf{92}, 26602 (2004).
\bibitem{Stiles-Single} M. D. Stiles, Jiang Xiao, A. Zangwill, Phys.
Rev. B \textbf{69}, 054408 (2004).
\bibitem{Sun-APL} J. Z. Sun \textit{et al.}, Appl. Phys. Lett.
\textbf{81}, 2202, (2002).
\bibitem{Sun3} J. Z. Sun \textit{et al.}, J. Appl. Phys. \textbf{93},
6859 (2003).
\bibitem{Barbaros- to be published} B.\"{O}zyilmaz
\textit{et al.}, to be published.
\bibitem{RalphScience} E. B. Myers \textit{et al}, Science
\textbf{285}, 867 (1999).
\bibitem{JiPRL} Y. Ji, C. L. Chien, and M. D. Stiles, Phys. Rev.
Lett. \textbf{90}, 106601 {2003}.
\bibitem{footnote1} In Ref.
\cite{JiPRL} the authors conclude that at high current densities
the single layer point contact experiments closely mimic trilayer
junctions in which excitations are caused by a spin polarized
current and detected via a GMR type of effect. Here the extended
film acts as a reference. Hence, excitations lead to a resistance
increase and show up as peaks in $dV/dI$.
\bibitem{Stiles private communications} M. D. Stiles, private
communications.
\bibitem{footnote2} The total height of the PD is fixed.
Therefore, an increase of the Co layer thickness reduces the
asymmetry in spin accumulation and hence could increase the
critical currents.
\end{thebibliography}

\end{document}